\documentclass[11pt]{article}
\usepackage{coling2018}
\usepackage{times}
\usepackage{url}
\usepackage{latexsym}

\usepackage{graphicx}
\usepackage[english]{babel}
\usepackage[T1]{fontenc}
\usepackage[utf8]{inputenc}
\usepackage{booktabs}
\usepackage[usenames, dvipsnames]{xcolor}
\usepackage{bm}
\usepackage{float}
\usepackage{makecell}
\usepackage{caption}
\usepackage{booktabs}
\usepackage{graphicx,wrapfig,lipsum}
\usepackage{wrapfig}
\usepackage{subfig}
\usepackage{natbib}
\usepackage{todonotes}
\usepackage{amsmath}
\usepackage{multirow}
\usepackage{xspace}

\usepackage{tabularx}
\usepackage{hyperref}

\newcommand{\Revised}[2]{\textcolor{orange}{#2}}

\definecolor{dgreen}{RGB}{34,139,34}

\newcommand{\FNC}{\textsc{fnc}\xspace}
\newcommand{\AGR}{\textsc{agr}\xspace}
\newcommand{\DSG}{\textsc{dsg}\xspace}
\newcommand{\DSC}{\textsc{dsc}\xspace}
\newcommand{\UNR}{\textsc{unr}\xspace}
\newcommand{\mF}{$F_1$m\xspace}

\title{A Retrospective Analysis of the Fake News Challenge \\ Stance Detection Task}

\author{Andreas Hanselowski$^\dagger$, Avinesh PVS$^\dagger$, Benjamin Schiller$^\dagger$, Felix Caspelherr$^\dagger$,\\ 
\textbf{Debanjan Chaudhuri$^\ddagger$\thanks{$^\star$The work by Debanjan Chaudhuri has been carried out during his internship at the Ubiquitous Knowledge Processing (UKP) Lab / Adaptive Preparation of Information from Heterogeneous Sources (AIPHES) from 01.01.2017 to 30.06.2017}, 
Christian M. Meyer$^\dagger$, Iryna Gurevych$^\dagger$}  \\
   Research Training Group AIPHES \\
  $^\dagger$Computer Science Department, Technische Universit\"at Darmstadt \\
   $^\ddagger$Smart Data Analytics, University of Bonn \\
\url{https://www.aiphes.tu-darmstadt.de} \\}

\date{}

\begin{document}

\maketitle
\begin{abstract}
The 2017 Fake News Challenge Stage 1 (FNC-1) shared task addressed a stance classification task as a crucial first step towards detecting fake news. To date, there is no in-depth analysis paper to critically discuss FNC-1’s experimental setup, reproduce the results, and draw conclusions for next-generation stance classification methods. In this paper, we provide such an in-depth analysis for the three top-performing systems. We first find that FNC-1’s proposed evaluation metric favors the majority class, which can be easily classified, and thus overestimates the true discriminative power of the methods. Therefore, we propose a new F1-based metric yielding a changed system ranking. Next, we compare the features and architectures used, which leads to a novel feature-rich stacked LSTM model that performs on par with the best systems, but is superior in predicting minority classes. To understand the methods’ ability to generalize, we derive a new dataset and perform both in-domain and cross-domain experiments. Our qualitative and quantitative study helps interpreting the original FNC-1 scores and understand which features help improving performance and why. Our new dataset and all source code used during the reproduction study are publicly available for future research\footnote{\url{https://github.com/UKPLab/coling2018_fake-news-challenge}}.

\end{abstract}

\section{Introduction}

Recently, \citet{fnc2017} organized the first Fake News Challenge\footnote{\url{http://www.fakenewschallenge.org/}} (FNC-1) in order to foster the development of AI technology to automatically detect fake news.
The challenge received much attention in the NLP community: 50 teams from both academia and industry participated.
The goal of the FNC-1 challenge is to determine the perspective (or \emph{stance}) of a news article relative to a given headline. 
An article's stance can either \emph{agree} or \emph{disagree} with the headline, \emph{discuss} the same topic, or it is completely \emph{unrelated}. Table~\ref{tbl:fnc_examples} shows four example documents illustrating these classes.

\blfootnote{
    \hspace{-0.65cm}  
    This work is licensed under a Creative Commons 
    Attribution 4.0 International License.
    License details:
    \url{http://creativecommons.org/licenses/by/4.0/}
}

Stance detection is a crucial building block for a variety of tasks, such as analyzing online debates \citep{walker2012stance, DBLP:conf/acl/SridharFHGW15, somasundaran2010recognizing}, determining the veracity of rumors on twitter \citep{lukasik2016hawkes, derczynski2017semeval}, or understanding the argumentative structure of persuasive essays \citep{stab2017parsing}. 
While stance detection has been previously focused on individual sentences\ or phrases, the systems participating in FNC-1 have to detect the stance of an entire document, which raises many new challenges. 
Although the disagreeing article of Table~\ref{tbl:fnc_examples} clearly leans against the headline's claim, the fourth sentence would agree to it if considered in isolation.

To properly learn from a scientific shared task, there are typically overview and analysis papers that compare the architectures, features, and results of the participating systems. To date, there is, however, no such paper for FNC-1, which is why we conduct a reproduction study of the top three participating systems.
Our goal is to independently verify the results reported in the challenge, which is an important asset in empirical research, to critically assess the experimental setup of FNC-1, and to learn building better methods by understanding their merits and drawbacks.
Based on our analysis of the shared task data, we first propose a new evaluation metric for FNC-1 and related document-level stance detection tasks, which is less affected by highly imbalanced datasets.
To understand the headroom of the state-of-the-art performance, we additionally estimate the upper bound for this task.
In a feature ablation study, we then identify which features contribute to solving the stance detection task. On the basis of our analysis, we combine ideas from previous systems and propose a novel architecture that performs on par with the state-of-the-art systems, but is better able to correctly classify difficult cases.
Since generalizability is crucial for the method's future impact, 
we finally introduce a new evaluation dataset and evaluate how well the FNC-1 models generalize to unseen data from a different domain. In addition to in-domain experiments, we also conduct cross-domain experiments in order to analyze the transfer potential of a method.

\newcommand\T{\rule{0pt}{2.6ex}}       
\newcommand\B{\rule[-1.2ex]{0pt}{0pt}} 

\begin{table*}[!t]
\small
\begin{tabular}{|p{1.4cm}|p{13.75cm}|}
\hline
\multicolumn{2}{|c|}{\textbf{Headline:} Hundreds of Palestinians flee floods in Gaza as Israel opens dams } \T \B\\
\hline
\T
\textbf{\textcolor{dgreen}{Agree (\AGR)}} & 
GAZA CITY (Ma'an) -- Hundreds of Palestinians were evacuated from their homes Sunday morning after Israeli authorities opened a number of dams near the border, flooding the Gaza Valley in the wake of a recent severe winter storm. The Gaza Ministry of Interior said in a statement that civil defense services and teams from the Ministry of Public Works had evacuated more than 80 families from both sides of the Gaza Valley (Wadi Gaza) after their homes flooded as water levels reached more than three meters [..] 
\B \\ \hline
\T
\textbf{\textcolor{blue}{Discuss (\DSC)}} & 
Palestinian officials say hundreds of Gazans were forced to evacuate after Israel opened the gates of several dams on the border with the Gaza Strip, and flooded at least 80 households. Israel has denied the claim as ``entirely false”. [..] \B \\ \hline
\T
\textbf{\textcolor{red}{Disagree (\DSG)}} &
Israel has rejected allegations by government officials in the Gaza strip that authorities were responsible for released storm waters flooding parts of the besieged area.
"The claim is entirely false, and southern Israel does not have any dams," said a statement from the Coordinator of Government Activities in the Territories (COGAT).
"Due to the recent rain, streams were flooded throughout the region with no connection to actions taken by the State of Israel."
At least 80 Palestinian families have been evacuated after water levels in the Gaza Valley (Wadi Gaza) rose to almost three meters. [..] \B\\ 
\hline 
\T
\textbf{\textcolor{orange}{Unrelated (\UNR)}} & Apple is continuing to experience `Hairgate' problems but they may just be a publicity stunt [..] \B \\
\hline
\end{tabular}
\caption{Headline and text snippets from document bodies with respective stances from the FNC dataset}
\vspace{-1.5ex}
\label{tbl:fnc_examples}
\end{table*}

\section{Related Work} \label{sec:RL}

Previous works in stance detection mostly considered target-specific stance prediction,
whereby the stance of a text entity with respect to a topic or a named entity is determined.
Target-specific stance prediction has been performed for
tweets \citep{mohammad2016semeval, AugensteinRVB16, zarrella2016mitre}\ and online debates \citep{walker2012stance,somasundaran2010recognizing, DBLP:conf/acl/SridharFHGW15}.
Such target-specific approaches are based on structural \citep{walker2012stance}, linguistic and lexical features \citep{somasundaran2010recognizing} and they jointly model disagreement only and collective stance using probabilistic soft logic \citep{DBLP:conf/acl/SridharFHGW15} or neural models \citep{zarrella2016mitre, DBLP:conf/ijcai/DuXHG17} with conditional encoding \citep{AugensteinRVB16}. 
Stance prediction in tweets \citep{mohammad2016semeval, AugensteinRVB16, DBLP:conf/ijcai/DuXHG17} and in online debates \citep{DBLP:conf/ijcnlp/HasanN13} is different from that of stance detection in a news article, which -- while similar -- is concerned with stance detection of a news article relative to a statement in natural language. 

To the best of our knowledge, there is yet no overview\ or analysis paper on FNC-1 similar to the shared task on detecting stance in twitter \citep{mohammad2016semeval, derczynski2017semeval, DBLP:conf/sepln/TauleMRRBP17}. To demonstrate the best scientific practices and achieve research transparency, we close this gap by systematically reviewing the  top-ranked systems at FNC-1.

The FNC-1 stance detection task is inspired by \citet{ferreira2016emergent}, who classify the stance of a single sentence of a news headline towards a specific claim. In FNC-1, however, the task is document-level stance detection, which requires the classification of an entire news article relative to a headline.
The top performing system in FNC-1 is called \emph{SOLAT in the SWEN} \citep{talos2017} by Talos Intelligence (henceforth: Talos). They use a combination of deep convolutional neural networks and gradient-boosted decision trees with lexical features. Team \emph{Athene} \citep{HanselowskiEtAl2017} won the second place with an ensemble of five multi-layer perceptrons (MLP) with six hidden layers each and handcrafted features. For the prediction they used hard voting. Finally, \emph{UCL Machine Reading} (\emph{UCLMR}) \citep{riedel2017simple} were placed third using a multi-layer perceptron with bag-of-words features.
Additionally, recently published work on FNC-1 use a two-step logistic regression based classifier \citep{bourgonje2017clickbait} and a stacked ensemble of five classifiers \citep{thorne2017fake} which achieve 9th and 11th places respectively. 
Although multiple systems\footnote{e.g., \url{http://web.stanford.edu/class/cs224n/reports.html}} participated at FNC-1, we focus on the top three systems in this paper, due to the availability of source code and our goal of analyzing what contributes most to good performance. In the remaining paper, we introduce and analyze these three systems in detail. 

\section{Reproduction of the Fake News Challenge FNC-1} 

In this section, we take a closer look at the challenge. We briefly discuss the task and dataset of FNC-1, describe the three top-ranked systems and reproduce their results.

\paragraph{FNC-1 task and dataset.} 
The task in FNC-1 is learning a classifier $f\colon (d, h) \mapsto s$ that predicts one of four stance labels $s \in S = \{\AGR, \DSG, \DSC, \UNR\}$ for a document $d$ with regard to a headline $h$. If headline and document cover different topics, the stance is $s = \UNR$ (unrelated). Otherwise, $s$ is \AGR if $d$ agrees and \DSG if $d$ disagrees with $h$. If $h$ and $d$ merely discuss the same topic, but $d$ does not take a definite position, $s$ will be \DSC.

To evaluate the challenge, the organizers provide a dataset\footnote{\url{https://github.com/FakeNewsChallenge/fnc-1-baseline}} of 300 topics. The topics are represented by claims with 5--20 news article documents each.
The dataset is derived from the Emergent project \citep{emergent2017} which addressed rumor debunking.
In the project, each news article document was summarized into a headline that reflects the stance of the whole document. 
Other than for rumor debunking, the FNC-1 organizers match each document with every summarized headline and then label the $(d, h)$ pair with one of the four stance labels $S$. 
To generate the unrelated class \UNR, headlines and documents belonging to different topics are randomly matched.
Document--headline pairs of 200 topics are reserved for training, the remaining document--headline pairs of 100 topics for testing. Topics, headlines, and documents are therefore not shared between the two data splits.
To prevent teams from using any unfair means by deriving the labels for the test set from the publicly available Emergent data,
the organizers additionally created 266 instances.
Table~\ref{tb:stat1} shows the corpus size and label distribution.

\begin{table*}
\centering
\begin{tabular}{l @{\qquad} c c c c @{\qquad} c c c c}
  \toprule
  Dataset & headlines & documents & tokens & instances  & \AGR & \DSG &  \DSC & \UNR  \\
  \midrule
  FNC-1     & 2,587 & 2,587 & 372 & 75,385 &  7.4\%   &	 2.0\%   &	17.7\%   &  72.8\% \\ 
  \bottomrule
\end{tabular}
\caption{Corpus statistics and label distribution for the FNC-1 dataset}
\vspace{-1.5ex}
\label{tb:stat1}
\end{table*}

\paragraph{Participating systems.}
For our reproduction study, we consider FNC-1's three top-ranked systems.
Talos Intelligence's SOLAT in the SWEN team \citep{talos2017} won the FNC-1 using their weighted average model (\emph{TalosComb}) of a deep convolutional neural network (\emph{TalosCNN}) and a gradient-boosted decision trees model (\emph{TalosTree}). 
\emph{TalosCNN} uses pre-trained word2vec embeddings\footnote{\label{fn:w2v}https://code.google.com/archive/p/word2vec/} passed through several convolutional layers followed by three fully-connected and a final softmax layer for classification.   
\emph{TalosTree} is based on word count, TF-IDF, sentiment, and singular-value decomposition features in combination with the word2vec embeddings.

Team \emph{Athene} \citep{HanselowskiEtAl2017} was ranked second.
They propose a multi-layer perceptron (MLP) inspired by the work of \citet{davisfake}. They extend the original model structure to six hidden and a softmax layer and they incorporate multiple hand-engineered features: unigrams, the cosine similarity of word embeddings of nouns and verbs between headline and document tokens, and topic models based on non-negative matrix factorization, latent Dirichlet allocation, and latent semantic indexing in addition to the baseline features provided by the FNC-1 organizers. 
Depending on the feature type, they either form separate feature vectors for document and headline, or a joint feature vector.

The UCL Machine Reading (\emph{UCLMR}) team propose an MLP as well, but use only a single hidden layer \citep{riedel2017simple}. Their system was ranked third.
As features, they use term frequency vectors of unigrams of the 5,000 most frequent words for the headlines and the documents. 
Additionally, they compute the cosine similarity between the TF-IDF vectors of the headline and document.
The resulting term frequency feature vectors of headline and document are concatenated along with the cosine similarity of the two TF-IDF vectors.

\paragraph{Reproduction.}
Following the instructions from the GitHub repositories of the three teams,\footnote{\raggedright \url{https://github.com/Cisco-Talos/fnc-1};\quad \url{https://github.com/hanselowski/athene_system};\quad \url{https://github.com/uclmr/fakenewschallenge}}
we could successfully reproduce the results reported in the competition without significant deviations. Table~\ref{tb:models} shows these results in the \FNC column of the \emph{FNC-FNC} setup, which means that the models were trained and tested on the FNC dataset. 
Since Talos use a combination of two models, we have also included the results of \emph{TalosCNN} and \emph{TalosTree}.
%
A first interesting finding is that \emph{TalosTree} even outperforms the combined model, since the CNN component performed poorly.
To understand the merits and drawbacks of the systems, we analyze the performance metrics and the features used, as discussed in the following sections.

\begin{table}[!tb]
\centering
\begin{tabular}{l@{\qquad}*{6}{@{\hspace{.30cm}}c}}
\toprule
\multirow{2}{*}{\bf Systems} & \multicolumn{6}{c}{\bf FNC-FNC} \\ 
& \FNC &\mF & \AGR & \DSG & \DSC & \UNR  \\ 
\midrule
Majority vote & .394 & .210 & 0.0 & 0.0 & 0.0 & .839  \\ 
TalosComb & .820 & .582 & \bf.539 & .035 & .760 & .994 \\
TalosTree & \bf.830 & .570 & .520 & .003 & .762 & .994 \\
TalosCNN & .502 & .308 & .258 & .092 &  0.0 & .882 \\
Athene    & .820 & .604 & .487 & .151 &  \bf.780 & \bf.996  \\
UCLMR     & .817 & .583 & .479  & .114 &  .747& .989  \\ 
featMLP   & .825 &  .607 & .530   & .151 & .766  & .982 \\ 
stackLSTM &  .821 &  \bf .609 & .501 & \bf.180 &  .757 &  .995 \\
Upper bound  & .859  & .754 & .588 & .667 & .765  & .997 \\ 
\bottomrule
\end{tabular}
\caption{\label{font-table} \FNC, \mF, and class-wise $F_1$ scores for the analyzed models on in-domain experiments}
\vspace{-2.5ex}
\label{tb:models}
\end{table}

\section{Performance evaluation} \label{sec:evaluation}

In this section, we critically assess the FNC-1 evaluation methodology and we determine a human upper bound for this task in order to identify the room for improvement for the document-level stance detection task.

\paragraph{Evaluation metrics.} \label{sec:eval_met}
The FNC-1 organizers propose the hierarchical evaluation metric \FNC, which first awards .25 points if a document is correctly classified as related (i.e., $s \in \{\AGR,\DSG,\DSC\}$) or \UNR to a given headline.
If it is related, .75 additional points are assigned if the model correctly classifies the document-headline pair as \AGR, \DSG, or \DSC.
The goal of this weighting schema is to balance out the large number of unrelated instances. 

Nevertheless, the metric fails to take into account the highly imbalanced class distribution of the three related classes \AGR, \DSG, and \DSC illustrated in Table~\ref{tb:stat1}.
Thus, models, which perform well on the majority class and poorly on the minority classes are favored. 
Since it is not difficult to separate related from unrelated instances (the best systems reach about $F_1 = .99$ for the \UNR class), a classifier that just randomly predicts one of the three related classes would already achieve a high \FNC score. A classifier that always predicts \DSC for the related documents even reaches \FNC = .833, which is even higher than the top-ranked system.

We therefore argue that the \FNC metric is not appropriate for validating the document-level stance detection task.
Instead, we propose the class-wise and the macro-averaged $F_1$ scores (\mF) as a new metric for this task that is not affected by the large size of the majority class.
The class-wise $F_1$ scores are the harmonic means of the precisions and recalls of the four classes, which are then averaged to the \mF metric.
The na\"ive approach of perfectly classifying \UNR and always predicting \DSC for the related classes, would achieve only \mF = .444, which is clearly different from the proposed systems.
By averaging over the individual classes' scores, \mF is also applicable to other datasets, which have a different class distribution than the FNC-1 dataset.
While the averaged \mF objectively reflects the quality of the prediction rather than the class distribution, we can also analyze which classes cannot be properly predicted yet. 

As the scores in Table~\ref{tb:models} indicate,
the performance of the three top-ranked systems reach only about \mF = .6. 
Our analysis reveals that \emph{TalosCNN} does not predict the \DSC class yielding an $F_1$ score of zero. Also the overall performance of this model is low and according to the \FNC metric, \emph{TalosTree} would even outperform \emph{TalosComb}. In contrast, \emph{TalosTree} returns almost no predictions for the \DSG class, although it performs exceptionally well in terms of \FNC.
This is because it often predicts the majority class \DSC for the related documents. Since there are only few \DSG instances in the dataset, the overall performance of this model appears high.


Considering the FNC-1 results according to our proposed \mF metric, the ranking of the three systems changes:
The \emph{TalosComb} and \emph{TalosTree} systems are slightly outperformed by \emph{UCLMR} and clearly outperformed by the Athene system.
This is because the two Talos models benefit from the \FNC metric definition, favoring the prediction of the majority classes \UNR and \DSC.
On smaller classes, such as \DSG, they perform much worse than Athene and \emph{UCLMR}. 
Using \mF as a metric, the Athene system would be ranked first, as it outperforms \emph{UCLMR} by 2.1 percentage points. In addition to that, Athene also works best on the \DSG, \DSC, and \UNR class.






\paragraph{Human upper bound.}
In addition to the issues with the evaluation metric, there is also no upper bound reported for the FNC-1 data, although this will help estimating the headroom of the proposed systems with regard to human performance.
Therefore, we ask five human raters to manually label 200 instances. 
The raters reach an overall inter-annotator agreement of Fleiss' $\kappa = .686$ \citep{fleiss1971measuring}, which is substantial and allows drawing tentative conclusions \citep{artstein2008}.
However, when ignoring the \UNR class, the inter-annotator agreement dramatically drops to $\kappa = .218$.
This indicates that differentiating between the three related classes \AGR, \DSG, and \DSC is difficult even for humans.

On the basis of the annotation, we also determine the most probable stance labels according to MACE \citep{hovy2013learning}, 
and compare them to the ground truth from the Emergent project.
The agreement of the labels in this case is better, 
reaching a Fleiss' $\kappa$ of .807 overall and .552 for the three related classes.
The MACE-based most probable label allows us to compute the human upper bound as \mF = .754, which we include in Table~\ref{tb:fnc-acr} along with the upper bound per class F1 scores \UNR = .997 \AGR = .588, \DSG =.667, and \DSC = .765.

\section{Analysis of models and features } 

In this section, we first perform an error analysis in order to be able to find out what the three best performing models are learning and in which cases they fail. In order to address the identified drawbacks, we conduct a systematic feature analysis and derive an alternative model based on our findings.  

\subsection{Error analysis } \label{sec:err}

Our error analysis for the three analyzed systems shows that the models fail in the following cases:
(1)~If there is lexical overlap between the headline and the document, 
the models classify the instance as one of the related classes, 
even in cases in which the two are unrelated.
(2)~If the document--headline pair is related, but only contains synonyms rather than the same tokens, the model often misclassifies the case as \UNR.
(3)~If keywords like \textit{reports}, \textit{said}, or \textit{allegedly} are detected, the systems classify the pair as \DSC.
(4)~The \DSG class is especially difficult to determine, as only few lexical indicators (e.g., \textit{false}, \textit{hoax}, \textit{fake}) are available as features. 
The disagreement is often expressed in complex terms which demands more sophisticated machine learning techniques. For example: ``If the bizarre story about\dots sounded outlandish, that's because it was''.
In appendix~\ref{sec:error}, we illustrate these errors with concrete examples.

The analysis shows that the models exploit the similarity between the headline and the document in terms of lexical overlap.
Lexical cue words, such as \textit{reports}, \textit{said}, \textit{false}, \textit{hoax} play an important role in classification. 
However, the systems fail when semantic relations between words need to be taken into account, complex negation instances are encountered, or the understanding of propositional content in general is required.
This is not surprising since the three models are based on $n$-grams, bag-of-words, topic models and lexicon-based features instead of capturing the semantics of the text.  
In this section, we test these features systematically and we propose new features and a new architecture for FNC-1.


\subsection{Feature analysis}\label{sec:fet}

\begin{figure*}[t]
\includegraphics[width=1.00\textwidth]{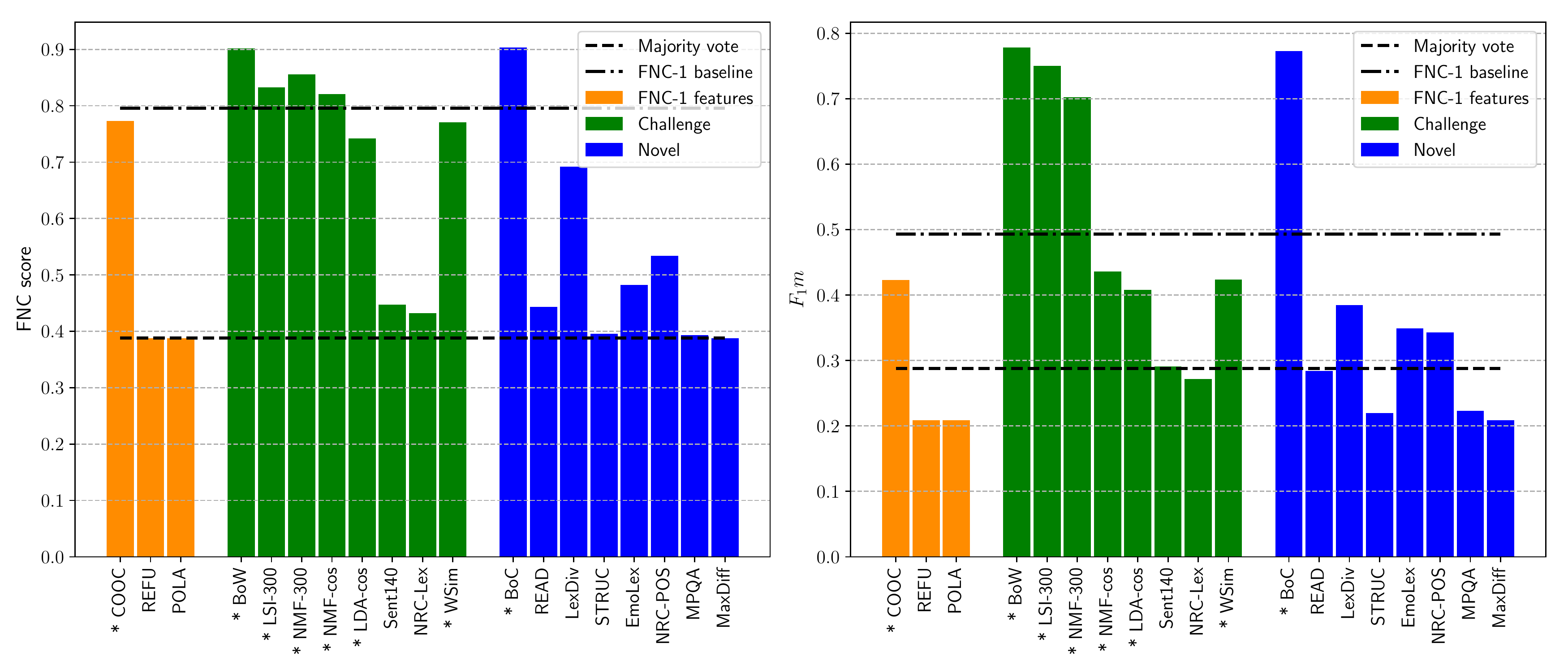}
\caption{Performance of the system based on individual features
}
\label{fig:feat_sel}
\vspace{-2.5ex}
\end{figure*}

Throughout our feature analysis, we use the Athene model, which performed best in terms of \mF and allows a large number of experiments due to its fast computation.  
All tests are performed on the FNC-1 development set with 10-fold cross-validation.
In the remaining section, we first discuss and evaluate the performance of each feature individually and then conduct an ablation test for groups of similar features. Detailed feature descriptions are included in the supplementary material (section~\ref{A:feat_detail}). Figure~\ref{fig:feat_sel} shows the system performance of the individual features discussed below. 

\paragraph{FNC-1 baseline features.} 
The FNC-1 organizers provide a gradient-boosting baseline using the co-occurrence (COOC) of word and character $n$-grams in the headline and the document as well as two lexicon-based features, which count the number of refuting (REFU) and polarity (POLA) words based on small word lists. 
Figure~\ref{fig:feat_sel} indicates that COOC performs well, whereas both lexicon-based features are on par with the majority vote baseline.

\paragraph{Challenge features.} 
The three analyzed FNC-1 systems rely on combinations of the following features:
Bag-of-words (BoW) unigram features, topic model features based on non-negative matrix factorization (NMF-300, NMF-cos) \citep{doi:10.1162/neco.2007.19.10.2756}, Latent Dirichlet Allocation (LDA-cos) \citep{DBLP:conf/nips/BleiNJ01}, Latent Semantic Indexing (LSI-300) \citep{deerwester1990indexing}, 
two lexicon-based features using 
NRC Hashtag Sentiment (NRC-Lex) and Sentiment140 (Sent140) \citep{MohammadKZ2013},
and word similarity features which measure the cosine similarity of pre-trained word2vec embeddings of nouns and verbs in the headlines and the documents (WSim). 
The topic models use 300 topics. Besides the concatenated topic vectors, we also consider the cosine similarity between the topics of document and headline (NMF-cos, LDA-cos). 
The BoW features perform best in terms of \mF.
While LSI-300, NMF-300 and NMF-cos topic models yield high scores, LDA-cos and WSim fall behind.

\paragraph{Novel features.}
We also analyze a number of novel features for the FNC-1 task which have not been used in the challenge.
Bag-of-character 3-grams (BoC) represent subword information. They show promising results in our setup. The structural features (STRUC) include the average word lengths of the headline and the document,
the number of paragraphs in the document and their average lengths. The low performance of this feature indicates that the structure of the headline and the documents is not indicative of their stance. 
Furthermore, we test readability features (READ) which estimate the complexity of a text. Less complex texts could be indicative of deficiently written fake news. We tried the following metrics for headline and document as a concatenated feature vector: SMOG grade \citep{mc1969smog}, Flesch-Kincaid grade level and Flesch reading ease \citep{kincaid1975derivation}, Gunning fog index \citep{vstajner2012can}, Coleman-Liau index \citep{coleman1975computer}, automated readability index \citep{senter1967automated}, LIX and RIX \citep{anderson1983lix}, McAlpine EFLAW Readability Score \citep{mcalpine1997global}, and Strain Index \citep{nirmaldasan-strain}.
However, in the present problem setting these features show only a low performance.
The same is true for the lexical diversity (LexDiv) metrics, type-token ratio, and the measure of textual diversity (MTLD) \citep{mccarthy2005assessment}.
We finally analyze the performance of features based on the following lexicons: MPQA \citep{Wilson:2005:RCP:1220575.1220619}, MaxDiff \citep{kiritchenko2014sentiment}, and EmoLex \citep{mohammad2010emotions}. These features are based on the sentiment, polarity, and emotion expressed by headlines and documents, which might be good indicators of an author's opinion. However, our results show that these lexicon-based features are not helpful. Even though the considered lexicons are important for fake-news detection (\cite{shu2017fake}, \cite{DBLP:journals/corr/HorneA17}), for stance detection, 
the properties captured by the lexicon-based features are not very useful.

\paragraph{Feature ablation test.}

We first remove all features that are more than 10\% below the FNC-1 baseline, since they mostly predict the majority class and thus harm the \mF score. 
In Figure~\ref{fig:feat_sel}, we mark these features with an asterisk (*). 
To quantify their contribution, we perform an ablation test across three groups of related features: (1) BoW and BoC (BoW/C), (2) LSI-topic, NMF-topic, NMF-cos, LDA-cos (Topic), and (3) NRC-POS and WSim (Oth).

Table~\ref{tb:ablation} show the
results of our ablation test. The BoW\ and BoC features have the biggest impact on the performance. While the topic models yield further improvements, the NRC-POS and WSim features are not helpful. 
Hence, we suggest BoW, BoC, and the four topic model based features as the most promising feature set.

We evaluate this feature set on the FNC-1 test dataset. The results are included in the \emph{featMLP} row of Table~\ref{tb:models} for the \emph{FNC-FNC} setting.
Although \emph{featMLP} with the revised feature selection outperforms the best performing FNC-1 system Athene in terms of \mF and \FNC score, the margin is not significant.
Similar to the three FNC-1 systems, we observe a .2 performance drop between the development and test dataset. This is most likely because of the 100 new topics in the test dataset, which have not been seen during training. Thus, the evaluation on the test set can be considered as an out-of-domain prediction.

\subsection{Model analysis}
In order to increase the overall performance, we conduct additional experiments with an ensemble of the three models \emph{featMLP}, \emph{TalosComb}, and \emph{UCLMR} using hard voting. 
However, we could not significantly improve the results.
Since all models struggle with the \DSG class, we have applied different under- and over-sampling techniques to balance the class distribution, 
but also this technique did not yield improved results.

\begin{table}[!b]
\vspace*{-2ex}
\begin{center}
\begin{tabular}{l @{\qquad} c c  @{\qquad} c @{\hspace{.25cm}} c c @{\qquad}  c @{\hspace{.2cm}} c c  @{\qquad} c c }
\toprule
   &\multicolumn{2}{c}{\textbf{Baselines}} & \multicolumn{3}{c}{\textbf{Only}} & \multicolumn{3}{c}{\textbf{All without}}\\
   & {Maj. vote} & {FNC-1} & {BoW/C} & {Topic} & {Oth} & {-BoW/C} & {-Topic} & {-Oth} & {All*} & {All} \\
\midrule
\AGR & 0.0  & .241 & \textbf{.772} & .637 &  0.0  & .665 & .714 & .722 & .713 & .675 \\
\DSG & 0.0 & .047 & .601 & .571 &  0.0 & .530 & .598 & \textbf{.616} & .573 & .455 \\
\DSC & 0.0 & .738 & .874 & .838 & .731 & .841 & .863 & \textbf{.87}6 & .870 & .835\\
\UNR & .835 & .970 & .991 & .983 & .964 & .982 & .989 & \textbf{.995} & .993 & .989\\
\midrule
\mF & .209 & .499 & .796 & .757 & .425 & .754 & .791 & \textbf{.802} & .787 & .738\\
\bottomrule
 \end{tabular}
\end{center}
\caption{Results of the feature ablation test. Baseline FNC-1 uses gradient boosting classifier with all FNC-1 baseline features. *\space states that only the preselected features are used (see Figure~\ref{fig:feat_sel}). }	
\vspace{-1.5ex}
    \label{tb:ablation}
\end{table}

\begin{figure*}[!h]
\centering
\includegraphics[width=.8\textwidth]{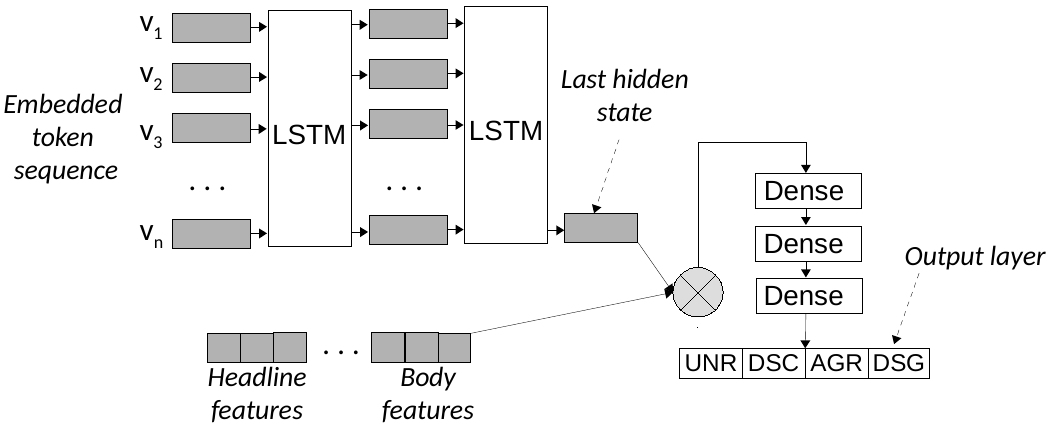}
\caption{Model Architecture of the feature-rich stackLSTM}
\label{fig:stackLSTM}
\vspace{-2.5ex}
\end{figure*}

In the error analysis, we observed that the feature-based systems lack semantic understanding.
Therefore, we combine a feature-based system with a model that is better able to capture the semantics based on word embeddings and sequential encoding. 
Sequential processing of information is important in order to get the meaning of the whole sentence, e.g. "It wasn’t long ago that Gary Bettman was ready to expand NHL." VS. "It was long ago that Gary Bettman wasn’t ready to expand NHL."
In Figure~\ref{fig:stackLSTM}, we introduce this \emph{stackLSTM} model, 
which combines the best feature set found in the ablation test with a stacked long short-term memory (LSTM) network \citep{hermans2013training}. We use 50-dimensional GloVe word embeddings\footnote{\url{http://nlp.stanford.edu/data/glove.twitter.27B.zip}} \citep{pennington2014glove} in order to generate sequences of word vectors of a headline--document\ pair. For this, we concatenate 
a maximum of 100 tokens of the headline and the document. 
These embedded word sequences $v_1, v_2, \dots, v_n$ are fed through two stacked LSTMs with a hidden state size of 100 with a dropout of 0.2 each. The last hidden state of the second LSTM is concatenated with the feature set and fed into a 3-layer neural network with 600 neurons each. Finally, we add a dense layer with four neurons and softmax activation function in order to retrieve the class probabilities. 

Table~\ref{tb:models} shows the performance of \emph{stackLSTM} for the \textit{FNC-FNC} setup.
Our model outperforms all other methods in terms of \mF. The difference to Athene and \emph{featMLP} is, however, not significant.
An important advantage of \emph{stackLSTM} is its improved performance for the \DSG class, which is the most difficult one to predict due to the low number of instances. This means that \emph{stackLSTM} correctly classifies a larger number of complex negation instances. 
The difference on this difficult \DSG class between \emph{stackLSTM} and all other methods is statistically significant (using Student's t-test).
The model predicts more often for the \DSG class and gets more of these examples correct without compromising the overall performance. One challenging \DSG example, which was correctly classified by the \emph{stackLSTM}, is given in Table~\ref{tbl:dsg_example}.
\newline



\begin{table*}[!t]
\small
\begin{tabular}{|p{15.5cm}|}
\hline
\T
\textbf{Headline:} NHL expansion ahead? No, says Gary Bettman \\
\hline
\T
\textbf{Article body:} It wasn’t very long ago that NHL commissioner Gary Bettman was treating talk of expansion as though he was being asked if he’d like an epidemic of Ebola. But recently the nature of the rhetoric has changed so much that the question is becoming not if, but when. ... \\
\hline
\end{tabular}
\caption{A correctly classified \DSG instance by the \emph{stackLSTM}}
\vspace{-1.5ex}
\label{tbl:dsg_example}
\end{table*}

\section{Analysis of the generalizability of the models}

To test the robustness of the models (i.e.\ how well they generalize to new datasets), we introduce novel test data for document-level stance detection based on the Argument Reasoning Comprehension (ARC) task\ proposed by \citet{habernal2017argument}. 
In this section, we describe the dataset, analyze the models' performance, and perform cross-domain experiments.

\paragraph{ARC dataset.}
\citet{habernal2017argument} manually select 188 debate topics with popular questions from the user debate section of the New York Times. 
For each topic, they collect user posts, which are highly ranked by other users, 
and create two claims representing two opposing views on the topic.  
Then, they ask crowd workers to decide whether a user post supports either of the two opposing claims or does not express a stance at all.
This Argument Reasoning Comprehension (ARC) dataset consists of typical controversial topics from the news domain, such as \textit{immigration}, \textit{schooling issues}, or \textit{international affairs}. While this is similar to the FNC-1 dataset, there are significant differences, as a user post is typically a multi-sentence statement\ representing one viewpoint on the topic. In contrast, the news articles of FNC-1 are longer and usually provide more balanced and detailed perspective on an issue. 

\begin{table*}
\centering
\begin{tabular}{l @{\qquad} c c c c @{\qquad} c c c c}
  \toprule
  Dataset & headlines & documents & tokens & instances  & \AGR & \DSG &  \DSC & \UNR  \\
  \midrule
  ARC       & 4,448 & 4,448 & 99  & 17,792 &  8.9\%   &	 10.0\%  &	6.1\%    &  75.0\% \\
  \bottomrule
\end{tabular}
\caption{Corpus statistics and label distribution for the ARC dataset}
\vspace{-1.5ex}
\label{tb:stat2}
\end{table*}

To allow using the ARC data for our FNC stance detection setup, we consider each user post as a document and
randomly select one of the two claims as the headline. We label the claim--document pair as \AGR if the claim has also been chosen by the workers, as \DSG if the workers chose the opposite claim, and as \DSC if the workers selected neither claim.
Table~\ref{tbl:arc_examples} shows an example of our revised ARC corpus structure.
In order to generate the unrelated instances, we randomly match the user posts with claims, but avoid that a user post is assigned to a claim from the same topic.
Table~\ref{tb:stat2} provides basic corpus statistics.
For training and testing, we split the corpus into 80\% training/validation set and 20\% testing set.

As for the FNC-1 corpus, we have also determined a human upper bound for the ARC dataset. 
Five subjects annotate 200 samples using the four classes.
Even though the overall Fleiss' $\kappa = .614$ is slightly lower compared to the FNC-1 corpus,
the agreement for the three related classes \AGR, \DSG, and \DSC is higher ($\kappa = .383$) than for FNC-1. 
The human upper bound based on MACE is \mF = .773. Table~\ref{tb:fnc-acr} contains also the class-wise $F_1$ scores.

\begin{table*}[!t]
\small
\begin{tabular}{|l|p{13.75cm}|}
\hline
\multicolumn{2}{|c|}{\bf Example from the original ARC dataset} \T \B\\
\hline
\T
\bf Topic & Do same-sex colleges play an important role in education, or are they outdated?
\B \\ \hline
\T
\bf User post  & Only 40 women's colleges are left in the U.S. And, while there are a variety of opinions on their value, to the women who have attended ... them, they have been ... tremendously valuable. ... \B \\ \hline
\T
\bf Claims  & \textbf{1.} Same-sex colleges are outdated \qquad \textbf{2.} Same-sex colleges are still relevant \B\\ 
\hline 
\T
\bf Label & Same-sex colleges are still relevant \B \\
\hline
\end{tabular}
\vspace{1ex}
\small
\begin{tabular}{|l|p{6.88cm}|p{6.84cm}|}
\hline
\multicolumn{3}{|c|}{\bf Generated instance in alignment with the FNC problem setting} \T \B\\
\hline
\T
\bf Stance  & \bf Headline & \bf Document
\B \\ \hline
\T
\AGR  & Same-sex colleges are still relevant & Only 40 women's colleges are left in the U.S. ...
\B \\ \hline
\end{tabular}
\caption{Example of the original ARC dataset and the generated instance to align with FNC dataset}
\vspace{-1ex}
\label{tbl:arc_examples}
\end{table*}

\paragraph{In-domain experiments ARC-ARC:}
The in-domain results for the ARC corpus listed in Table~\ref{tb:models2} show that the overall performance of all models decreases. 
Since the models have been constructed to perform well on the FNC-1 dataset, this is not surprising. 
Nevertheless, for the ARC corpus, the models are better able to distinguish between \AGR and \DSG instances.
We assume this is because the number of \DSG instances is substantially larger and is similar to the number \AGR instances.
The classification of the \DSC instances, on the other hand, turns out to be more challenging on the ARC corpus.
This is because even if a user post is related to the claim, it often does not explicitly refer to it.
%
With \emph{TalosComb} being best, the Talos models were better able to generalize to the new data.
Even though the \emph{stackLSTM} is again better on the more difficult minority class (in this case \DSC), 
the structure and features of \emph{TalosComb} seem to be more appropriate for this problem setting.

\begin{table}[!tb]
\centering
\begin{tabular}{l@{\qquad}*{6}{@{\hspace{.30cm}}c}}
\toprule
\multirow{2}{*}{\bf Systems} & \multicolumn{6}{c}{\bf ARC-ARC} \\ 
& \FNC & \mF & \AGR & \DSG & \DSC & \UNR \\ 
\midrule
Majority vote  & .430 & .214  &  0.0 & 0.0 & 0.0 & .857 \\ 
TalosComb & \bf .725 & \bf .573 & \bf.593   & \bf.598  & .160  & \bf.944\\
Athene   &  .680  & .548    & .516 &  .482 & .190  & .933 \\
UCLMR     & .667    & .519 & .517   & .503 & .121  & .932\\ 
featMLP   & .690   & .526 & .526   & .506 & .144  & .934  \\ 
stackLSTM & .685 & .524 & .451 & .518 & .\bf 194 & .935 \\
Upper bound & .796 & .773 & .710  & .857 & .571  & .954  \\ 
\bottomrule
\end{tabular}
\caption{\label{font-table} \FNC, \mF, and class-wise $F_1$ scores for the analyzed models on in-domain experiments}
\vspace{-1.5ex}
\label{tb:models2}
\end{table}

\paragraph{Cross-domain experiments:}
In the cross-domain setting we train on the training data of one corpus and evaluate on the test data of the other corpus. 
The experiments in Table~\ref{tb:fnc-acr} show that the performance of the models is substantially better than the majority vote baseline.
We therefore conclude that the two problem settings are related and exhibit a common structure.
The results suggest that \emph{TalosComb} is best able to learn from the ARC corpus, as it is also superior in the \emph{ARC-FNC} setting. 
The \emph{stackLSTM}, on the other hand, yields best results when trained on the FNC corpus as the \emph{FNC-ARC} setting suggests.

\begin{table}[!tb]
\centering
\begin{tabular}{l@{\qquad}*{6}{@{\hspace{.30cm}}c}@{\hspace{.60cm}}*{6}{@{\hspace{.30cm}}c}}
\toprule
\multirow{2}{*}{\bf Systems} & \multicolumn{6}{c}{\bf FNC-ARC} & \multicolumn{6}{c}{\bf ARC-FNC} \\ 
& \FNC &\mF & \AGR & \DSG & \DSC & \UNR  & \FNC & \mF & \AGR & \DSG & \DSC & \UNR \\ 
\midrule
Majority vote & .430 & .214  &  0.0 & 0.0 & 0.0 & .857 & .394 & .210 & 0.0  & 0.0 & 0.0  & .839  \\ 
TalosComb & .584 & .365 & .336   & 0.0 & .195  & \bf.929 & .607 & \bf .388 & .279  & \bf.183 & .\bf113  & \bf.977\\
Athene    & .523 & .340 & \bf .340 & \bf .244 &  .138 & .894  &  .548 & .321 & .277 &  .097 & .028 & .882 \\
UCLMR     & .557 & .358  & .271   & .064 & \bf.201  & .896  & .482 & .288 & .234 & .109 & .080  & .728\\ 
featMLP   & .586 &  .389 & .321   & .159 & .171  & .906 & .585 & .351  & .322 & .111 & .033  & .939 \\ 
stackLSTM &  \bf.591 &  \bf .401 & .321 & .191 &  .182 &  .910 &  \bf .613 & .373 & \bf .343 & .116 &  .082 & .950\\
Upper bound        & .796         & .773 & .710  & .857 & .571  & .954 & .859  & .754 & .588 & .667 & .765  & .997 \\ 
\bottomrule
\end{tabular}
\caption{\label{font-table} FNC, \mF and class-wise $F_1$ scores(\mF) based on cross-domain experiments}
\vspace{-2.5ex}
\label{tb:fnc-acr}
\end{table}

\section{Discussion and conclusion}

In this paper, we conducted a thorough analysis of the Fake News Challenge stage one stance detection task.
Although this is common for shared tasks, there is yet no analysis or reproduction study of this task, which is why we close this gap.
Given that the challenge has attracted much attention in the NLP community with 50 participating teams, a detailed analysis is valuable as it provides insights into the problem setting and lessons learned for upcoming competitions.   
In our investigation, we evaluated the performance of the three top-scoring systems, critically assessed the experimental setup, and performed a detailed feature analysis, in which we identify high-performing features for the task yielding a new model.
We conducted an error analysis and found that the models mostly rely on the lexical overlap for classification. 
To assess how well the models generalize to a similar problem setting, we experimented with a second, newly derived corpus.  
We also propose a new evaluation metric based on $F_1$ scores, since the challenge's metric is highly affected by the imbalanced class distribution of the test data.
Using this evaluation setup, the ranking of the top three systems changes.  
Based on our analysis, we conclude that the investigated stance detection problem is challenging, 
since the best performing features are not yet able to resolve the difficult cases. 
Thus, more sophisticated machine learning techniques are needed, which have a deeper semantic understanding, and are able to determine the stance on the basis of propositional content instead of relying on lexical features.

\section{Acknowledgements}

This work has been supported by the German Research Foundation as part of the Research
Training Group “Adaptive Preparation of Information from Heterogeneous Sources” (AIPHES) at the Technische Universit\"at Darmstadt under grant No. GRK 1994/1.

\bibliographystyle{acl_natbib}
\bibliography{coling2018}

\appendix

\section{Supplemental Material}

\subsection{Features: Detailed description}\label{A:feat_detail}
\begin{description}
\item [BoW/BoC features] We use bag-of-words (BoW) 1- and 2-grams with 5,000 tokens vocabulary for the headline as well as the document. For the BoW feature, based on a technique by \citet{das2007yahoo}, we add a negation tag "\_NEG" as prefix to every word between special negation keywords (e.g. "not", "never", "no") until the next punctuation mark appears. For the bag-of-characters (BoC) 3-grams are chosen with 5,000 tokens vocabulary, too. For the BoW/BoC feature we use the TF to extract the vocabulary and to build the feature vectors of headline and document. The resulting TF vectors of headline and document get concatenated afterwards. Feature \textit{co-occurrence} (FNC-1 baseline feature) counts how many times word 1-/2-/4-grams, character 2-/4-/8-/16-grams, and stop words of the headline appear in the first 100, first 255 characters of the document, and how often they appear in the document overall.
\item [Topic models] We use non-negative matrix factorization (NMF) \citep{doi:10.1162/neco.2007.19.10.2756}, latent semantic indexing (LSI) \citep{deerwester1990indexing}, and latent Dirichlet allocation (LDA) \citep{DBLP:conf/nips/BleiNJ01} to create topic models out of which we create independent features. For each topic model, we extract 300 topics out of the headline and document texts. Afterwards, we compute the similarity of headlines and bodies to the found topics separately and either concatenate the feature vectors (NMF, LSI) or calculate the cosine distance between them as a single valued feature (NMF, LDA).
\item [Lexicon-based features] These features are based on the NRC Hashtag Sentiment and Sentiment140 lexicon \citep{kiritchenko2014sentiment, MohammadKZ2013, zhu2014nrc}, as well as for the MPQA lexicon \citep{Wilson:2005:RCP:1220575.1220619} and MaxDiff Twitter lexicon \citep{rosenthal-EtAl:2015:SemEval, kiritchenko2014sentiment}. All named lexicons hold values that signal the sentiment/polarity for each word. The features are computed separately for headline and document, and constructed as proposed by \citet{MohammadKZ2013}: First, we count how many words with positive, negative, and without polarity are found in the text. Two features sum up the positive and negative polarity values of the words in the texts and another two features are set by finding the word with the maximum positive and negative polarity value in the text. Finally, the last word in the text with negative or positive polarity is taken as a feature. Since the MaxDiff Twitter lexicon also contains 2-grams, we decide to take them into account as well, whereas for the other lexicons only 1-grams incorporated. Additionally, we base features on the EmoLex lexicon \citep{mohammad2010emotions, Mohammad13}. For all its words, it holds up to eight emotions (anger, fear, anticipation, trust, surprise, sadness, joy, disgust), based on the context they frequently appear in. For headline and document respectively, the emotions for all words are counted as a feature vector. The resulting vectors for headline and document are then concatenated. Lastly, the baseline features \textit{polarity words} and \textit{refuting words} are added. The first one counts refuting words (e.g. "fake", "hoax"), divides the sum by two, and takes the remainder as a feature signaling the polarity of headline or document. The latter one sets a binary feature for each refuting word (e.g. "fraud", "deny") appearing in the headline or document.
\item [Readability features] We measure the readability of headline and document with SMOG grade (only document), Flesch-Kincaid grade level, Flesch reading ease, and Gunning fog index \citep{vstajner2012can}, Coleman-Liau index \citep{coleman1975computer}, automated readability index \citep{senter1967automated}, LIX and RIX \citep{anderson1983lix}, McAlpine EFLAW Readability Score \citep{mcalpine1997global}, Strain Index \citep{nirmaldasan-strain}. The SMOG grade is only valid if a text has at least 30 sentences, and thus is only implemented for the bodies. 
\item [Lexical features] As lexical features we implement the type-token-ratio (TTR) and the measure of textual lexical diversity (MTLD) \citep{mccarthy2005assessment} for the document, and only type-token-ratio for the headline, since MTLD needs at least 50 tokens to be valid. Also, the baseline feature \textit{word overlap} belongs to this group. It divides the cardinality of the intersection of unique words in headline and document by the cardinality of the union of unique words in headline and document.
\item [POS features] The POS features amongst others include counters for nouns, personal pronouns, verbs and verbs in past tense, adverbs, nouns and proper nouns, cardinal numbers, punctuations, the ratio of quoted words, and also the frequency of the three least common words in the text. The headline feature also contains a value for the percentage of stop words and the number of verb phrases, which showed good results in the work of \citet{DBLP:journals/corr/HorneA17}. For the \textit{\Revised{Stanford }{}word-similarity} feature, \Revised{}[{which are mainly based on \citet{ferreira2016emergent}} we calculated average word embeddings (pre-trained word2vec model\footnote{https://code.google.com/archive/p/word2vec/}) for all verbs (retrieved with Stanford Core NLP toolkit\footnote{https://stanfordnlp.github.io/CoreNLP/}) of headline/document separately. The cosine similarity between the averaged embeddings of headline and document is taken as a feature, as well as the hungarian distance between verbs of headline and document based on the paraphrase database\footnote{http://www.cis.upenn.edu/~ccb/ppdb/}. The same computation is done for all nouns of headline and document. 
Additionally the average sentiment of the headline and the average sentiment of the document is used as a feature. A count of negating words of the headline and the document is added to the feature vector as well as the distance from the negated word to the root of the sentence. The number of average words per sentence of headline and document is another feature. The aforementioned features are improved by only selecting a predefined number of sentences of document and headline. Therefore the sentences are ordered by TF-IDF score.
\item [Structural features] The structural features contain the average word length of the headline and document, and the number of paragraphs and average paragraph length of the document.
\end{description}

\subsection{Misclassified examples identified in the error analysis} \label{sec:error}

Example 1.  \newline
(ground truth: "unrelated", system predicts: "agree")\\
Headline:        CNN: Doctor Took Mid-Surgery Selfie with Unconscious Joan Rivers\\
        
\noindent Document:        "A TEENAGER woke up during brain surgery to ask doctors how it was going.
        Iga Jasica, 19, was having an op to remove a tumour at when the anaesthetic wore off and she struck up a conversation with the medics still working on her."\\
    
\noindent Example 2. \newline
(ground truth: "agree", system predicts: "unrelated")\\
Headline: Three Boobs Are Most Likely Two Boobs and a Lie	\\
 
\noindent Document: The woman who claimed she had a third breast has been proved a hoax.\\
    
\noindent Example 3. \newline
(ground truth: "disagree", system predicts: "discuss")\\
        Headline: Woman pays 20,000 for third breast to make herself LESS attractive to men\\
        
\noindent Document: The woman who reported that she added a third breast was most likely lying.\\

\noindent Example 4. \newline
    (ground truth: "disagree", system predicts: "agree")\\
        Headline: Disgusting! Joan Rivers Doc Gwen Korovin’s Sick Selfie EXPOSED — Last Photo Of Comic Icon, When She Was Under Anesthesia\\
        
\noindent Document: If the bizarre story about Joan Rivers' doctor pausing to take a "selfie" in the operating room minutes before the 81-year-old comedienne went into cardiac arrest on August 29 sounded outlandish, that's because it was.

\end{document}